\documentclass[12pt,preprint]{aastex}
\accepted{21 Jan. 2009}

\slugcomment{To appear in the Astrophysical Journal}

\lefthead{Skinner et al.}
\righthead{V1735 Cygni}

\begin{document}

\title{X-ray Emission from the FU Orionis Star V1735 Cygni}

\author{Stephen L. Skinner and Kimberly R. Sokal}
\affil{Center for Astrophysics and Space Astronomy (CASA), 
Univ. of Colorado, Boulder, CO 80309-0389 (skinners@origins.colorado.edu) }

\author{Manuel G\"{u}del and Kevin R. Briggs}
\affil{Institute of Astronomy, ETH Zurich, Wolfgang-Pauli-Str. 27,
 8093 Zurich, Switzerland}

%
\newcommand{\ltsimeq}{\raisebox{-0.6ex}{$\,\stackrel{\raisebox{-.2ex}%
{$\textstyle<$}}{\sim}\,$}}
%
\newcommand{\gtsimeq}{\raisebox{-0.6ex}{$\,\stackrel{\raisebox{-.2ex}%
{$\textstyle>$}}{\sim}\,$}}

\begin{abstract}
The variable star V1735 Cyg (= Elias 1-12) lies in the 
IC 5146 dark cloud and
is a member of the class of FU Orionis objects whose dramatic optical
brightenings are thought to be linked to episodic accretion.
We report the first X-ray detections of V1735 Cyg and a
deeply-embedded  class I protostar lying 24$''$ to its northeast. 
X-ray spectra obtained with  EPIC  on {\em XMM-Newton} 
reveal very high-temperature plasma (kT $>$ 5 keV) in both
objects, but no large flares. Such hard X-ray emission is not anticipated 
from accretion shocks and is a signature of magnetic processes.
We place these new results into the context of what is
presently known about the X-ray properties of FU Orionis
stars and other accreting young stellar objects.
\end{abstract}




%

\section{Introduction}
IC 5146 originally  referred to the bright nebula 
surrounding the B1V star BD $+$46$^{\circ}$3474
and was already identified in early observations by 
Espin (1900). The nebula is commonly referred to
as the  Cocoon Nebula. 
Wolf (1904) noted that faint stars were present in
the nebula but were much less numerous immediately 
outside it. Later optical studies by Walker (1959),
Herbig (1960), and Forte \& Orsatti (1984) confirmed
that a cluster of more than 100 faint young stars surrounds
BD $+$46$^{\circ}$3474.  Herbig \& Dahm (2002) showed that
many of these faint objects exhibit H$\alpha$ emission and
are pre-main sequence (PMS)
T Tauri stars. They obtained a cluster distance 
d = 1200 pc (with an uncertainty of at least $\pm$180 pc), 
mean extinction A$_{V}$ = 3.0 $\pm$ 0.2 mag,
and age $\sim$1 Myr. 

Dark cloud filaments extend 
$\approx$2$^{\circ}$ westward from BD $+$46$^{\circ}$3474
(Wolf 1904), and this region is commonly referred to as
the IC 5146 dark cloud. 
Deep near-IR (HK) imaging of a portion of
the  filament known as the Northern Streamer
by Lada, Alves, \& Lada (1999) revealed heavily-embedded 
young stars with extinctions  up to A$_{V}$ $\approx$ 20 - 50 mag.  
Submillimeter surveys have detected  dense cores in some
of the  filaments that may be prestellar (Kramer et al. 2003),
suggesting that stars are actively forming within the 
dark cloud. 

A  recent {\em Spitzer} study of the IC 5146 star-forming
region using the IRAC and MIPS cameras was presented by
Harvey et al. (2008; hereafter H08). They reexamined the 
distance to the region based on a comparison with the
Orion  Nebula Cluster (ONC), which is of similar age.
H08 concluded that: (i) the most likely distance to 
the optical cluster surrounding BD $+$46$^{\circ}$3474 
is 950 $\pm$ 80 pc, and (ii) the IC 5146 dark cloud 
lying to the west of BD $+$46$^{\circ}$3474 is at a
similar distance. 
H08 used infrared colors to identify more than one hundred
candidate young stellar objects (YSOs) in IC 5146 and modeled
their spectral energy distributions to infer disk properties.
They concluded that  $\approx$40\% of the cluster members
detected by {\em Spitzer} are surrounded by disks.

One of the most unusual objects identified to date 
in the IC 5146 dark cloud is the  variable star 
V1735 Cyg, lying $\approx$1$^{\circ}$ west of
BD $+$46$^{\circ}$3474. It  has an infrared spectral
energy distribution characteristic of a
class II
\footnote{The YSO class is determined from the 
slope $\alpha$ of the infrared spectral 
energy distribution, where 
$\alpha$ = $d$log($\lambda F_{\lambda}$)/$d$log $\lambda$.
Greene et al. (1994) used flux densities $F_{\lambda}$ in the
2.2 - 10 $\mu$m range to compute $\alpha$ but
H08 used data over the range $\lambda$ = 2.2 - 24 $\mu$m
where available.  Greene et al. (1994) classify
sources with $\alpha$ $>$ 0.3 as class I,
0.3 $>$ $\alpha$ $\geq$ $-$0.3 as flat-spectrum,
$-$0.3 $>$ $\alpha$ $\geq$ $-$1.6 as class II, 
and  $\alpha$ $<$ $-$1.6 as class III.}
source (H08).
It was discovered in the near-IR survey of
Elias  (1978) and is object number 12 (Elias 1-12)
in his catalog. On the basis of a comparison with previous
red Palomar plates, Elias  concluded that this object 
brightened by at least 5 magnitudes at R  sometime 
between  1952 and 1965. He also noted that
its spectrum is similar to the eruptive variable
FU Orionis and he classified V1735 Cyg as an 
FU Orionis star (FUor). These rare objects have undergone
strong optical outbursts that are thought to be
related to episodic disk accretion, as reviewed 
by Hartmann \& Kenyon (1996; hereafter HK96). 
The enhanced accretion
gives rise to a strong wind and CO observations 
of V1735 Cyg provide evidence for mass outflow
(Levreault 1983). The radio continuum
emission detected from V1735 Cyg is consistent with free-free
wind emission at an ionized mass-loss rate 
$\dot{\rm M}$ $\sim$ 10$^{-7}$ M$_{\odot}$ yr$^{-1}$
(Rodriguez \& Hartmann 1992). Cold dust is also present
near the star as evidenced by a submillimeter
detection (Sandell \& Weintraub 2001, hereafter SW01).

We present here the results of an X-ray observation
of the IC 5146 dark cloud  obtained with {\em XMM-Newton},
centered near  V1735 Cyg. This observation was obtained 
as part of a broader program aimed at determining if FUors are 
X-ray sources and establishing  their X-ray properties,
which prior to this survey were unknown. We recently
reported the X-ray detection of the prototype star
FU Orionis with {\em XMM-Newton}, which shows an unusual 
double-absorption X-ray spectrum with a soft and hard component 
(Skinner, Briggs, \& G\"{u}del  2006; hereafter SBG06).
We show here that V1735 Cyg is also an X-ray source
with a hard component.  We also report
the discovery of a second heavily-absorbed X-ray source 
located $\approx$24$''$ northeast of V1735 Cyg that is 
associated with a class I protostar detected in the
mid-infrared by {\em Spitzer}.

\section{XMM-Newton Observations}

The observation  began on 
2006 July 22 at 15:27 UT and ended at
23:15 UT. Data were acquired with the European Photon
Imaging Camera (EPIC), which provides charge-coupled
device (CCD) 
imaging spectroscopy from the pn camera
(Str\"{u}der et al. 2001) and two nearly
identical MOS cameras (MOS1 and MOS2;
Turner et al. 2001). The observation was
obtained in full-window mode using the medium
optical blocking filter. The EPIC cameras
provide  energy coverage in the range
E $\approx$ 0.2 - 15 keV with energy 
resolution E/$\Delta$E $\approx$ 20 - 50.
The MOS cameras provide the best on-axis angular 
resolution with FWHM $\approx$ 4.3$''$ at
1.5 keV.

Data were reduced using the {\em XMM-Newton}
Science Analysis System (SAS vers. 7.1).
To insure that the latest calibration was 
applied, we ran the SAS pipeline processing
tasks $epchain$ and $emchain$ to generate updated 
pn and MOS event files. These event files 
were then filtered to select good
event patterns. 
Inspection of pn background light curves showed 
two short intervals of elevated background,
and these intervals were excluded during spectral
extraction. We obtained 27,275 s of total PN exposure
and 26,000 s of usable exposure after removing the 
two high background
intervals. No time-filtering of the MOS data was
required and the usable exposure times were
31,663 s (MOS1) and 31,667 s (MOS2).

Spectra and
light curves were extracted from  circular
regions  of radius  R$_{e}$ = 15$''$ centered
on V1735 Cyg and other sources of interest, 
corresponding to $\approx$68\%
encircled energy at 1.5 keV. Background
spectra and light curves were obtained
from circular source-free regions near
the source. The SAS tasks
{\em rmfgen} and {\em arfgen} were used to
generate source-specific response matrix
files (RMFs) and auxiliary response files
(ARFs) for spectral analysis. The data were
analyzed using the {\em XANADU} software package
\footnote{The {\em XANADU} X-ray analysis software package
is developed and maintained by NASA's High Energy
Astrophysics Science Archive Research Center. See
http://heasarc.gsfc.nasa.gov/docs/xanadu/xanadu.html
for further information.},
including {\em XSPEC} vers. 12.4.0.

\section{Results}

\subsection{X-ray Properties of V1735 Cygni}

Figure 1  shows the {\em XMM-Newton} MOS image near
V1735 Cyg. Two X-ray sources separated by $\approx$ 24$''$ are
present. The position of the southwest source (Table 1)
is offset by 0.$''$91 from the {\em HST}
Guide Star Catalog (GSC v2.3.2) optical position of V1735 Cyg
(HST J214720.66$+$473203.8), 1.$''$02 from 
its {\em 2MASS} position (2MASS J214720.65$+$473203.5;
Fig. 2), and 0.$''$83 from its {\em Spitzer} position
(Spitzer J214720.65$+$473203.8;   H08; Fig. 3). These offsets
are nearly identical to the mean rms positional offset of
1.0$''$ determined by comparing the X-ray
and near-IR positions of 18 sources in 
the EPIC field having  {\em 2MASS} counterparts.
These offsets are consistent with the expected
{\em XMM-Newton} positional accuracy of $\leq$1$''$ 
at the signal-to-noise (S/N) ratio of our data
\footnote{XMM-Newton EPIC calibration data can be found at:~ \\
http://xmm2.esac.esa.int/docs/documents/CAL-TN-0018.pdf}.
We find no other optical or near-IR sources 
within 10$''$ of V1735 Cyg in the {\em HST} GSC,
USNO B1, {\em 2MASS}, or  HEASARC on-line extragalactic catalogs
\footnote{http://heasarc.gsfc.nasa.gov/docs/archive.html}
and therefore conclude
that V1735 Cyg is the most likely counterpart to
X-ray source XMM J214720.58$+$473204.3.

The JCMT submillimeter position of V1735 Cyg given by SW01
is offset 4.$''$6 - 4.$''$8 northeast of  the {\em HST} GSC,
{\em Spitzer}, and {\em XMM-Newton} positions of V1735 Cyg.
A similar northeasterly offset is noted for the JCMT position of
the submillimeter source SM1 (Sec. 3.2). This may indicate a 
slight systematic northeasterly shift of JCMT positions
relative to optical, IR (H08), and X-ray (Table 1) positions   
but it should be kept in mind that the JCMT observations 
were sensitive to cold dust emission which does not always  
peak at the stellar position.

Figure 4 shows the X-ray light curve of V1735 Cyg
from the pn detector.  No large amplitude variability
is detected. A $\chi^2$ variability test on the pn light
curve binned at 2000 s intervals gives a  
probability of constant count rate  P$_{const}$ = 0.81
($\chi^2$/dof = 8.5/13) and a larger binsize of
5000 s gives  P$_{const}$ = 0.56. 
Analysis of the individual MOS
light curves binned at 5000 s intervals give 
P$_{const}$ = 0.46 (MOS1) and  P$_{const}$ = 0.36 (MOS2). Thus,
no statistically significant X-ray variability is found
for V1735 Cyg.

Figure 5 shows the pn X-ray spectrum of V1735 Cyg. 
Faint line features are visible near 1.86 keV (Si XIII)
and 6.67 keV (Fe K$\alpha$ complex, including Fe XXV), 
implying  thermal emission rather than a power-law spectrum.
We have thus fitted the spectra with  single-temperature
(1T) and two-temperature (2T) $apec$ optically thin
plasma models (Table 2). The 1T $apec$ model is
statistically acceptable and requires a high
but uncertain plasma temperature with a lower
bound kT $\geq$ 6.4 keV (90\% confidence).
There are insufficient counts in the spectra to
place a useful upper bound on the plasma temperature.

The 1T $apec$ model gives an absorption
column density equivalent to visual extinction
A$_{V}$ = 2.9 [2.1 - 4.8] mag using the 
conversion N$_{\rm H}$ = 2.2 $\times$ 10$^{21}$ A$_{V}$
~cm$^{-2}$ mag$^{-1}$ (Gorenstein (1975), and a similar  
value using the  Ryter (1996) conversion. Alternatively, the
conversion  of Vuong et al. (2003) from data for the $\rho$ Oph
dark cloud is
N$_{\rm H}$ = 1.6 $\times$ 10$^{21}$ A$_{V}$ ~cm$^{-2}$ mag$^{-1}$
(assuming A$_{J}$/A$_{V}$ = 0.28)
and gives   A$_{V}$ = 4.1 [2.9 - 6.6] mag.
The above  values are less than the values
A$_{V}$ $\approx$ 8.0 - 10.8  mag determined from infrared
studies of V1735 Cyg (Elias 1978; H08). Assuming
that the infrared values are correct, this could be
an indirect clue that the 1T $apec$ model
is an oversimplification. 

A 2T $apec$ model gives a value of the  $\chi^2$ fit statistic
that is nearly identical to that of 1T $apec$, but converges
to a higher absorption that equates to 
A$_{V}$ = 6.5 [1.8 - 10.6] mag (Gorenstein 1975;
Table 2 notes). This is in 
better  agreement with the above IR estimates but the 
temperatures of the cool (kT$_{1}$ $<$ 1 keV) and hot 
(kT$_{2}$ $>$ 5 keV) components in the 2T model are 
poorly-constrained. Any cool plasma would be quite
heavily-absorbed and difficult to detect, but could
be present since most low-mass YSOs do show both a cool 
and hot component. This is generally true for T Tauri stars 
(Preibisch et al. 2005; hereafter P05) and is also the case
for FU Ori (SBG06). 

To summarize the above: the simplest
1T $apec$ thermal model that provides an acceptable
fit to the V1735 Cyg X-ray spectrum gives an equivalent
A$_{\rm V}$ that is a factor of $\sim$2 less than published values
based on IR data. This discrepancy can be alleviated by
adding a second cool plasma component to the X-ray
spectral model (2T $apec$). There is good reason to suspect
that such a cool component may be present, but acceptable
X-ray fits can be obtained without it.

\subsection{X-ray Properties of SM1-X}

The X-ray source detected $\approx$24$''$ northeast of
V1735 Cyg is not visible in 2MASS near-IR images
(Fig. 2), but is clearly seen in  {\em Spitzer}
IRAC images (Fig. 3) and is also detected by
MIPS at 24 and 70  microns (H08). The {\em Spitzer} 
counterpart to the X-ray source is listed as source 
number 34 in the catalog of H08 who identified it 
with the JCMT submillimeter source SM1 (SW01). The JCMT
position of SM1 is offset 3.$''$5 - 3.$''$8
northeast of the  {\em Spitzer} and {\em XMM}
positions, similar to the JCMT offset noted for
V1735 Cyg (Sec. 3.1). Assuming that some (or all)
of the JCMT positional offset is systematic, it 
seems  likely that the submillimeter source is 
the same object detected by  {\em Spitzer} and
{\em XMM-Newton}. We thus hereafter refer to this
X-ray source as SM1-X (= XMM J214722.71$+$473215.3; 
Table 1).  Using  {\em Spitzer} data, H08 concluded that this 
object is an extremely cold, embedded class I source.
Figure 3 also reveals two other faint nearby
mid-IR sources listed as class I objects 32 and 
33 in H08. Their 8 $\mu$m flux densities are 
more than an order of magnitude lower than SM1 and V1735 Cyg,
and neither was detected in X-rays.

The pn X-ray light curve of SM1-X (Fig. 4) does not
show any large impulsive flares but there is  
a slow decline in count rate by a factor of 
$\sim$2 during the first 20 ksec of the observation.
A  $\chi^2$ analysis of the pn light curve binned at 
2000 s intervals gives P$_{const}$ = 0.19 and a 
larger binsize of 3000 s gives P$_{const}$ = 0.03.
Thus, slow low-level X-ray variability is likely 
present in SM1-X.

The pn spectrum (Fig. 5)  shows a hard component
including a  detection of the  Fe K emission 
line complex at 6.67 keV. There is no clear detection
of a fluorescent Fe line at 6.4 keV but there is some
weak emission redward of  the Fe K line visible in
unbinned spectra and higher
S/N spectra would be useful to determine
if a fluorescent line is present.

A 1T $apec$ model is acceptable
(Table 2) and gives a best-fit temperature 
kT = 6.5 [3.9 - 11.9; 90\% confidence] keV.
The absorption column density is much higher than that of
V1735 Cyg and yields an equivalent visual extinction
A$_{V}$ = 35 [29 - 42] mag (Gorenstein 1975).
This X-ray source is thus very heavily obscured,
as is typical of class I objects. Since there are
no obvious emission lines in the spectrum apart 
from the strong Fe K line, we also tried fitting
the data with a power-law model plus a Gaussian
line component to reproduce the Fe K line. This
model is acceptable (Table 2) and gives a best-fit 
line energy E$_{line}$ = 6.64 [6.51 - 6.77] keV,
in accord with the Fe K identification,
and a reduced $\chi^2$ that is identical to the 
1T $apec$ model. The absorption inferred from the 
power-law model is about 10\% larger than 
determined from $apec$.

\vspace*{0.5in}

\section{Discussion}

\subsection{X-ray Emission from FU Orionis Stars}

FUors were originally identified on the basis of 
their dramatic optical outbursts (Herbig 1966, 1977).
Only a few of these {\em classical} FUors are known
(HK96, SW01). We have now observed the four classical
FUors that define the class
with {\em XMM-Newton} as part of our exploratory
survey aimed at determining their X-ray properties.
As summarized in Table 3, we have detected V1735 Cyg 
and the prototype FU Ori, but V1057 Cyg and V1551 Cyg
were undetected.

Both V1735 Cyg and FU Ori show high-temperature plasma
at kT$_{hot}$ $>$ 5 keV that is characteristic of 
magnetic (e.g. coronal)  processes 
and incompatible with accretion shocks. This is substantiated by
a definite detection of  the Fe K line at 6.67 keV 
in FU Ori, and a possible detection in V1735 Cyg (Fig. 5).
This line complex emits maximum power
at log T$_{max}$ $\approx$ 7.6 K. Moreover, this 
high-temperature plasma is detected even in the
absence of any obvious large flares, suggesting
that its presence is not transient.  In the case
of FU Ori, the hot X-ray  component is viewed under
much higher absorption N$_{\rm H}$ $\sim$ 10$^{23}$ cm$^{-2}$
(SBG06) than is  anticipated from visual extinction estimates
(Table 3). The origin of this excess absorption is not 
yet known but either the disk, cold accreting gas,
or FU Ori's strong wind may contribute (SBG06).

A cool plasma component is also detected in FU Ori (Fig. 5)
at a best-fit temperature kT$_{cool}$ = 
0.7 [0.1 - 1.0; 90\% conf.] keV, and the absorption associated
with this component is consistent with A$_{\rm V}$ estimates
(SBG06; Table 3). There is no clear detection of a cool plasma 
component in V1735 Cyg, but 
the absorption inferred from X-ray spectral fits is in
better agreement with IR values if a cool component is 
included in the spectral model (Sec. 3.1).

One of the primary reasons for undertaking an
X-ray survey of FUors was to determine if
accretion shocks might play a role in their
X-ray emission. They are accreting at very
high rates up to $\dot{M}_{acc}$ $\sim$10$^{-4}$ 
M$_{\odot}$ yr$^{-1}$ (HK96). 
An accretion shock is expected to reveal itself
in X-rays as soft emission at a characteristic 
temperature
kT$_{shock}$ $\approx$ 0.02v$_{100}^2$  keV
where v$_{100}$ is the infall speed in units
of 100 km s$^{-1}$ (Ulrich 1976). For plausible 
infall speeds of a few hundred km s$^{-1}$, 
the expected shock temperature
is  kT$_{shock}$ $\sim$ 0.1 - 0.2 keV.

Accretion shock emission may be present in the
classical T Tauri star (CTTS) TW Hya ($\dot{M}_{acc}$ 
$\sim$ 10$^{-8}$ M$_{\odot}$ yr$^{-1}$),
based on the detection of cool plasma for which
a high density was inferred from line flux ratios
in X-ray grating observations (Kastner
et al. 2002; Stelzer \& Schmitt 2004). 
No X-ray grating observations have yet been
obtained of FUors but, as noted above, a cool plasma  
component is detected in FU Ori and may also be
present in V1735 Cyg. Even so, the temperature
of the cool component kT = 0.7 [0.1 - 1.0] keV
(T $\approx$ 8 MK) for FU Ori is
several times higher than expected for an accretion 
shock. Plasma at T  $\approx$ 8 MK is quite
typical of T Tauri stars and is also frequently
detected in coronally-active stars that are
not accreting (P05). It is thus by no means certain 
that  the cool emission seen in the FU Ori 
X-ray spectrum is accretion-related.

\subsubsection{X-ray Constraints on Stellar Luminosity and Mass}

The X-ray luminosities of both V1735 Cyg and FU Ori
(Table 3; Fig. 6) are at the high end of the range observed
for other YSOs (excluding massive hot stars) such as 
those in the ONC (P05),
Serpens (Giardino et al. 2007, hereafter G07), and
the Taurus molecular cloud (G\"{u}del et al. 2007; 
Telleschi et al. 2007, hereafter T07). The high
L$_{\rm X}$ is particularly intriguing because it
is present even in the absence of any obvious large
flares. This could be an indication that the 
central stars are at the high  end of the
luminosity (or mass) range typically found for T Tauri stars,  
since it is known that L$_{\rm X}$ increases with 
stellar luminosity (L$_{star}$) and stellar mass
(M$_{star}$)  in T Tauri stars (P05; T07).
Interestingly, it has been suggested that the central star
in V1735 Cyg is a high-luminosity object (Elias 1978)
and a similar claim has been made for FU Ori 
(Herbig, Petrov, \& Duemmler 2003; Petrov \& Herbig 2008).

In general, the spectral types, masses,  and luminosities  
of the central stars in FUors are not well-determined 
because of the difficulty in disentangling spectral 
features of the star from those of the luminous accretion 
disk, and imprecise knowledge of disk properties and
disk inclination angle  
(e.g. Kenyon, Hartmann, \& Hewett 1988; HK96).
Because of the high accretion rates of FUors,
the accretion-disk luminosity during outbursts
can be comparable to
that of the underlying star and will contribute significantly
to the bolometric luminosity 
L$_{bol}$ $=$ L$_{star}$ $+$ L$_{disk}$ 
given in Table 3.

Given the uncertainties in stellar properties of FUors,
it is worthwhile to seek  constraints on 
L$_{star}$ or M$_{star}$ using the X-ray data.
To obtain a quantitative estimate of L$_{star}$
for V1735 Cyg, we assume that the underlying star
is a single object resembling  a TTS and
that its L$_{\rm X}$ was not  temporarily
elevated during our observation because of a flare. 
We use the correlation
between L$_{star}$ and L$_{\rm X}$ found for
CTTS in the XEST survey of Taurus, which has
less scatter than the COUP ONC sample.
The linear regression fit for the XEST CTTS sample 
is (T07; Fig. 6):
log L$_{\rm X}$ = 1.16 ($\pm$0.09) log (L$_{star}$/L$_{\odot}$)
 $+$ 29.83 ($\pm$0.06) ergs s$^{-1}$. 
Using  L$_{\rm X}$ from Table 1, this relation gives
L$_{star}$ $\sim$ 5 - 22 L$_{\odot}$ for V1735 Cyg. 
This estimate accounts for uncertainties in 
L$_{\rm X}$ for V1735 Cyg from different spectral
models (Table 2) and uncertainties in the
XEST regression fit. But, other factors such as age
differences between the Taurus CTTS sample and
V1735 Cyg could also affect the comparison.

The inferred luminosity  L$_{star}$ $\sim$ 5 - 22 L$_{\odot}$
is reasonable for  a TTS or even a class I object,
but is at the high end of their ranges 
(Doppmann et al. 2005; Getman et al. 2005;
P05; T07).  L$_{\rm X}$ is also correlated with
stellar mass and the XEST regression fit using
the bisector method for CTTS in Taurus gives (T07):
log L$_{\rm X}$ = 1.98 ($\pm$0.20) log (M$_{star}$/M$_{\odot}$)
$+$ 30.24 ($\pm$0.06) ergs s$^{-1}$.
This  would require
M$_{star}$ $\gtsimeq$ 1.7 M$_{\odot}$ to account
for the X-ray luminosity of V1735 Cyg.
A similar lower limit on M$_{star}$ is obtained
from the COUP ONC regression fits (P05).
A reliable upper limit on  M$_{star}$ cannot be 
obtained from the XEST or COUP regression fits 
because they are based on stars with
M$_{star}$ $\ltsimeq$ 2 M$_{\odot}$.

The above estimates assume that the underlying star
in V1735 Cyg is a T Tauri-like object and that it
obeys the regression relations for TTS. This is a 
reasonable assumption given that the optical spectrum
of at least one FUor, V1057 Cyg, resembled a TTS
prior to outburst (Herbig 1977). Unfortunately, little
is known about the pre-outburst nature of V1735 Cyg
(Elias 1978). The above estimates also assume that
the X-ray emission of V1735 Cyg arises in a single 
star and not an unresolved binary. Obviously, if 
the X-ray luminosity is the summed contribution of
two stars in a close binary system then the above 
values of L$_{star}$ and M$_{star}$ would be overestimated. 
However, we are not aware of any reports so far that 
V1735 Cyg is a spectroscopic binary and high-resolution
Keck I spectra have shown that the  radial velocity of 
FU Ori is constant to within $\pm$0.3 km s$^{-1}$ over 
three years (Petrov \& Herbig 2008).

The preceding analysis shows that the X-ray luminosity
of V1735 Cyg is consistent with the value expected 
for a central star that is a relatively luminous
(L$_{star}$ $\sim$ 5 - 22 L$_{\odot}$) or 
massive (M$_{star}$ $\gtsimeq$ 1.7 M$_{\odot}$)
single T Tauri star. By analogy, a similar conclusion 
holds for FU Ori since its X-ray luminosity is 
comparable to that of V1735 Cyg (Table 3).
If future observations demonstrate that either
L$_{star}$ or M$_{star}$ is significantly less
than these values, we would be forced to
conclude that L$_{\rm X}$ in these X-ray bright
FUors is enhanced above the normal levels 
found for T Tauri stars. Whether any dramatic
increase in L$_{\rm X}$ might have occurred during their
previous outbursts is not  known because there
are no pre-outburst X-ray data for classical FUors.

Following the above approach, upper limits
on L$_{star}$ and M$_{star}$ can be obtained for
the undetected FUors V1057 Cyg and V1515 Cyg.
Applying the XEST CTTS regression
results to the tighter upper limit
log L$_{\rm X}$ $\leq$ 30.0 ergs s$^{-1}$ for
V1057 Cyg (Table 3) gives
L$_{star}$ $\leq$ 1.4 ($\pm$0.2) L$_{\odot}$
and M$_{star}$  $\leq$ 0.8 ($\pm$0.3) M$_{\odot}$.
Although the upper limits on L$_{\rm X}$
for V1057 Cyg and V1515 Cyg are well below
the values of the two detected FUors, they are 
still  not particularly stringent. The
majority of YSOs in nearby star-forming regions
such as the ONC and  Taurus molecular cloud
emit at levels log L$_{\rm X}$ $\leq$ 30.0 
ergs s$^{-1}$ (Getman et al. 2005; P05; T07).
Models of the mid- to far-IR emission of V1057 Cyg 
and V1515 Cyg suggest that they are viewed
nearly pole-on through evacuated cavities in surrounding
envelopes  (Kenyon \& Hartmann 1991; Millan-Gabet et al. 2006;
Zhu et al. 2008). If that is the case, then
a low-extinction path would be available for 
stellar X-rays to escape. Thus, 
deeper observations of V1057 Cyg and 
V1515 Cyg are  needed to search for fainter
emission that could be present if the central
stars  are lower mass  objects of a few tenths
of a solar mass.

\subsection{X-ray Emission from Class I Protostars}

As a result of deep surveys of star-forming regions 
like COUP (Getman et al. 2005) and XEST (G\"{u}del et al. 2007), 
a wealth of X-ray data now exists for class II (CTTS) and class III
(weak-lined TTS) YSOs. Much less is known about the X-ray
properties of heavily-obscured class I protostars, but 
some trends are now beginning to emerge.
For example, there is good evidence
that X-ray absorption column densities decrease  in progressing from
embedded class I sources to  optically-revealed non-accreting 
class III objects (G07; Prisinzano et al. 2008, hereafter
P08). Values of N$_{\rm H}$ $\sim$
10$^{22}$ - 10$^{23}$ cm$^{-2}$ are commonly found for
class I objects. There is a trend for {\em lower}
unabsorbed X-ray luminosity in class I sources, as compared to
class II and III (G07; P08), for reasons that are
not yet fully understood. Most class I sources have
X-ray luminosities  log L$_{\rm X}$ $<$ 31.0 ergs s$^{-1}$
(P05; G07; Ozawa, Grosso, \& Montmerle 2005; T07) but in 
rare cases higher. Also, sources with higher unabsorbed
L$_{\rm X}$ tend to have higher average plasma 
temperatures (P08).
Class I sources are often variable in X-rays (P08), and 
some dramatic magnetic-reconnection flares were detected
in class I sources during a long {\em Chandra} exposure
of the $\rho$ Oph molecular cloud (Imanishi, Koyama, \&
Tsuboi 2001; hereafter I01). Interestingly, some class I 
sources in $\rho$ Oph were detected {\em only} when they flared.
Thus, the two class I objects near
V1735 Cyg (sources 32 and 33 in Fig. 3) which were undetected
in our {\em XMM} observation may well be capable of transient
X-ray outbursts.

The properties of the class I object SM1-X analyzed in this
work are in general agreement with the above results, but
its high L$_{\rm X}$  is exceptional (Table 1).
The high plasma temperature of SM1-X is consistent with
the trend noted above for higher average temperatures
in sources with high  L$_{\rm X}$.
The high plasma temperature of SM1-X determined from the 
thermal spectral model (Table 2)  
points clearly to a magnetic origin for its X-ray emission. 
Temperatures derived from
spectral fits of heavily-absorbed objects like SM1-X can be
biased toward high values because the low-energy part of
the spectrum is masked, but the presence of the Fe K
line leaves little doubt that hot X-ray plasma is present.
Because of the high X-ray absorption, any soft emission
that might be present due to accretion shocks, shocked outflows, or 
a cool coronal component escapes detection.

The high X-ray  luminosity log L$_{\rm X}$ = 31.6 
ergs s$^{-1}$ inferred for SM1-X (Table 1) is remarkable
because it was observed even in the absence of any large
flares. Comparable X-ray luminosities have been reported
in some class I objects such as YLW 15A in $\rho$ Oph 
(I01), but were recorded during flares. However, as we
have noted, there is a slow decline in the X-ray
count rate of SM1-X during the first $\sim$20 ks
of the observation (Fig. 4), and it is conceivable
that this represents the decay tail of a large flare
that peaked prior to the start of the {\em XMM} observation.

There are other processes  besides flare decay that
could be responsible for the slow X-ray variability of
SM1-X. Similar slow variability has also been seen in
other X-ray sources such as  IRS 7B (= IRS 7 X$_{\rm E}$)
in the Corona Australis
dark cloud (Hamaguchi et al. 2005, 2006).
This X-ray source corresponds to a young class 0/I
transitional object (Groppi et al. 2007).
One possibility is that the slow variability is 
tied to stellar rotation. That is, part of the
X-ray emitting region may be periodically occulted
by the star or disk.
Such modulation has been detected in some ONC
YSOs (Flaccomio et al. 2005). 
Longer time monitoring of specific class I 
protostars over timescales of days to weeks would be 
useful to search for X-ray periodicity.

The presence of large-amplitude X-ray flares
in some class I sources and the existence
of very hot plasma provide strong (but indirect)
evidence that these objects do have 
magnetic fields. These fields are evidently
present in very young objects since the 
estimated age of the low-mass pre-main sequence  
population in IC 5146 is $\sim$0.6 Myr 
(H08). This poses the interesting
theoretical question of how young class I objects
which are presumably fully-convective  can
produce magnetic fields.  Are the fields 
internally-generated by dynamo action, as 
simulations suggest could be the case for
fully-convective M dwarfs (Browning 2008)?
Or, are the fields primordial and  inherited
from the parent molecular cloud?

Lastly, it is worth calling attention to the 
striking  similarity between the X-ray spectrum
of the class I source SM1-X and the 
hard (kT $>$ 2 keV) spectral component of 
FU Ori (Fig. 5). Thermal models give similar
high temperatures kT $\approx$ 6 - 7 keV and
high absorption column densities 
N$_{\rm H}$ $\sim$ 10$^{23}$ cm$^{-2}$ 
in both cases (Table 2; SBG06).
Some subtle differences are present
however. For example, the weak fluorescent Fe
line at 6.4 keV that is likely present in FU Ori is
not clearly detected in SM1-X. But, the fluorescent Fe line
has been seen in other class I sources 
(I01) and is usually interpreted as arising in
cold near-neutral material near the star - possibly
the disk (Tsujimoto et al. 2005). The differences
seen between SM1-X and FU Ori at low energies 
below 2 keV may be absorption-related and do not
necessarily reflect intrinsic differences in their underlying
spectral properties. The strong similarity in
their hard spectral component suggests a possible link
between FU Ori and class I protostars. 
Similarities between FUors and
class I objects have also been noted at 
submillimeter wavelengths (SW01).

\section{Summary and Open Questions}

Our exploratory survey  with {\em XMM-Newton}
has obtained the first
pointed X-ray observations of four classical
FUors. The prototype FU Ori as well as V1735 Cyg
were detected, but V1057 Cyg and V1515 Cyg were not.
Both detections show high-temperature X-ray 
plasma that is undoubtedly associated with 
magnetic (e.g. coronal) processes, but no large-amplitude
flares. Cool X-ray plasma (kT$_{cool}$ =
0.7 [0.1 - 1.0] keV) is also present in FU Ori, 
but not clearly detected in V1735 Cyg. The cool plasma 
temperature is at least a factor of $\sim$2 higher 
than expected for an accretion shock, so a magnetic 
(coronal) origin seems likely. But, the temperature 
of the cool component is quite uncertain and 
a pending deep {\em Chandra} observation may provide
tighter temperature constraints.

The X-ray luminosities of both FU Ori and V1735 Cyg
are at the high-end of the range observed for 
low-mass YSOs.  This fact, along with the known
correlation between L$_{\rm X}$ and stellar luminosity
(or mass) in YSOs, suggests that the central stars in
FU Ori and V1735 Cyg are high-luminosity objects,
as has previously been proposed. Thus, X-ray observations
offer the potential to determine properties of the central
star in FUor star$+$disk systems that are exceedingly
difficult to extract from their complex optical and
infrared spectra.

We  have obtained the first  X-ray detection of the newly-discovered
class I protostar SM1-X, located near V1735 Cyg. The X-ray
spectrum is heavily-absorbed with an equivalent visual
extinction A$_{\rm V}$ $\sim$ 35 mag, confirming that 
SM1-X is still deeply-embedded and  very young.
Its high X-ray temperature is characteristic of 
magnetic processes and adds to the accumulating
evidence that class I sources have
magnetic fields. Whether these fields are internally-generated
or primordial is still an open question. The origin of the 
slow low-level X-ray variabilility seen in SM1-X and 
other class I objects such as IRS 7B in CrA is still a
mystery, but it may be linked to stellar rotation.
If this can be confirmed from longer  monitoring
over timescales of $\sim$days to weeks,
then X-ray observations may be able to provide information
on rotation periods for embedded class I protostars that is
difficult to obtain by other methods.

The X-ray spectrum of the class I object SM1-X is 
remarkably similar to the hard component detected
in FU Ori. General similarities between FUors and
class I sources have also been noted at submillimeter 
wavelengths (SW01). This suggests a possible link 
between FU Ori and class I protostars. The precise
nature of any such relationship is unclear, but
it may be that FU Ori is now in the process of emerging
from the deeply-embedded class I phase.

\acknowledgements

This research was supported by NASA  grants 
NNX06AE93G and NNG05GE69G. We thank T. Preibisch
for information on statistical properties of 
COUP X-ray sources and P. Harvey for details
of {\em Spitzer} results.  This work 
is based on observations obtained with 
{\em XMM-Newton}, an ESA science mission with
instruments and contributions directly funded
by ESA member states and the USA (NASA).
This research utilized data obtained
through the High Energy Astrophysics Science
Archive Research Center (HEASARC) Online
Service provided by the NASA Goddard Space 
Flight Center. This work made use of data 
products from the Two Micron All Sky Survey
(2MASS), which is a joint project of the Univ. 
of Massachusetts and the Infrared Processing
and Analysis Center/California Institute of 
Technology (CalTech),
funded by NASA and NSF. This work is based in part
on archival data obtained with the {\em Spitzer} Space
Telescope, which is operated by the Jet Propulsion
Laboratory, CalTech, under a contract with NASA.

\clearpage

\begin{deluxetable}{lll}
\tabletypesize{\scriptsize}
\tablewidth{0pc}
\tablecaption{X-ray Source Properties} 
\tablehead{
\colhead{       }      &
\colhead{V1735 Cyg}        &
\colhead{SM1-X} 
}
\startdata
R.A. (h,m,s)                    & 21 47 20.585         & 21 47 22.710  \nl
Decl. ($^{\circ}$, $'$, $''$)   & $+$47 32 04.3        & $+$47 32 15.3 \nl
pn rate (c/ks)          & 5.3 $\pm$ 1.3        & 8.1 $\pm$ 2.2 \nl
MOS rate (c/ks)         & 2.4 $\pm$ 0.7        & 2.9 $\pm$ 0.6 \nl
log L$_{\rm X}$ (ergs s$^{-1}$)\tablenotemark{a}   & 31.0 $\pm$ 0.2   & 31.6 $\pm$ 0.1    \nl
log (L$_{\rm X}$/L$_{bol}$)\tablenotemark{b}       & $-$4.95          & ...              \nl

\enddata
\tablecomments{
The J2000.0 X-ray positions are based on the {\em XMM-Newton}
standard processing source list and have been corrected for 
systematic offsets using the SAS processing task $eposcorr$.
The mean count rates ($\pm$1$\sigma$)  are background-subtracted in the 0.5 - 7 keV
range, computed from light curves binned at 3000 s (PN) and 5000 s (per MOS). 
The MOS count rates are per MOS. The total exposure times 
were 27,275 s (PN), 31,663 s (MOS1), and 31,667 s (MOS2).}
\tablenotetext{a}{L$_{\rm X}$ (0.5 - 7 keV) is unabsorbed at d = 1 kpc and the 
uncertainties in L$_{\rm X}$ reflect the range of values
obtained for different emission models (Table 2).}
\tablenotetext{b}{We adopt L$_{bol}$ = 235 L$_{\odot}$ (H08) for V1735 Cyg,
which includes the mid-IR excess  from the disk. 
There is no reliable L$_{bol}$ estimate for SM1-X.}

\end{deluxetable}

\begin{deluxetable}{llll}
\tabletypesize{\scriptsize}
\tablewidth{0pc}
\tablecaption{{\em XMM-Newton} Spectral Fit Results 
   \label{tbl-1}}
\tablehead{
\colhead{Parameter}      &
\colhead{       }        &
\colhead{       } 
}
\startdata
Object                            & V1735 Cyg                               &  SM1X               & SM1X         \nl
Model                             & 1T APEC\tablenotemark{a}                &  1T APEC            & PL $+$ line\tablenotemark{b}        \nl
N$_{\rm H}$ (10$^{22}$ cm$^{-2}$) & 0.65 [0.46 - 1.06]                      &  7.82 [6.53 - 9.24] & 8.86 [6.68 - 13.1]        \nl
kT$_{1}$ (keV)                    & 13.7 [6.37 - ...]                       &  6.48 [3.87 - 11.9] & ...        \nl
norm$_{1}$ (10$^{-5}$)            & 3.28 [2.77 - 4.39]                      &  16.9 [13.3 - 25.3] & ...        \nl 
Z (Z$_{\odot}$)                   & \{1.0\}\tablenotemark{c}   & \{1.0\}\tablenotemark{d}         & ...        \nl
$\Gamma_{pl}$                     & ...                                     & ...                 & 2.08 [1.54 - 2.59]  \nl
norm$_{pl}$ (10$^{-5}$)           & ...                                     & ...                 & 9.86 [3.54 - 60.3]  \nl
$\chi^2$/dof                      & 23.1/24                                 & 31.5/35             & 28.7/32         \nl
$\chi^2_{red}$                    & 0.96                                    & 0.90                & 0.90          \nl
F$_{\rm X}$ (10$^{-14}$ ergs cm$^{-2}$ s$^{-1}$) & 3.95 (5.32)              & 9.46 (27.7)         & 9.22 (41.0)      \nl
log L$_{\rm X}$ (ergs s$^{-1}$)   & 30.80                                   & 31.52               & 31.69        \nl
\enddata
\tablecomments{
Tabulated values are based on  simultaneous fits of all three EPIC  spectra (PN, MOS1, MOS2)
using the APEC optically thin plasma model and the power law (PL) model
in XSPEC v12.4.0. 
The spectra were rebinned to a minimum of
15 (PN) and 10 (MOS) counts per bin. The tabulated parameters
are absorption column density (N$_{\rm H}$), plasma temperature (kT),
XSPEC normalization (norm), abundance (Z), and
photon power-law index ($\Gamma$). 
For the APEC models, the XSPEC norm is related to the 
emission measure (EM) by EM = 4$\pi$10$^{14}$d$_{cm}^2$$\times$norm, 
where d$_{cm}$ is the stellar distance in cm. Curly braces enclosing
the abundance mean that it was held fixed at solar
abundances referenced to Anders \& Grevesse (1989).
Brackets enclose 90\% confidence intervals.
X-ray flux (F$_{\rm X}$) is the  observed (absorbed) value followed
in parentheses by the unabsorbed value in the 0.5 - 7.0 keV range.
X-ray luminosity (L$_{\rm X}$) is the unabsorbed value in the
0.5 - 7.0 keV range. 
A distance of  1 kpc is assumed (Herbig \& Dahm 2002; Harvey et. al 2008).} 

\tablenotetext{a}{A 2T APEC model gives a larger N$_{\rm H}$ = 1.43 [0.39 - 2.34] 
  $\times$ 10$^{22}$ cm$^{-2}$ and a larger unabsorbed flux
  F$_{\rm X}$ = 1.33 $\times$ 10$^{-13}$ ergs cm$^{-2}$ s$^{-1}$
  (log L$_{\rm X}$ = 31.20 ergs s$^{-1}$), with  a
  nearly identical value of   $\chi^2_{red}$ = 0.94. 
  But the temperatures of the cool (kT$_{1}$ $<$ 1 keV) and hot 
  (kT$_{1}$ $>$ 5 keV) components   are poorly-constrained. }
\tablenotetext{b}{The power-law (PL) model includes a Gaussian line component
 to model the Fe K emission line. It converged to a best-fit line energy 
  E$_{line}$ = 6.64 [6.51 - 6.77] keV, 
norm$_{line}$ = 1.11 [0.42  - 2.15] $\times$ 10$^{-6}$, and 
$\sigma_{line}$ = 67 [0 - 168] eV, where $\sigma_{line}$ = FWHM/2.35.}
\tablenotetext{c}{Allowing the abundance to vary gives Z = 0.85 Z$_{\odot}$,
                  with no significant change in N$_{\rm H}$ or $\chi^2_{red}$,
                  but higher kT. The 90\% confidence bounds on Z are not tightly 
                  constrained by   the data.}
\tablenotetext{d}{Allowing the abundance to vary gives 
                  Z = 0.55 [0.23 - 1.01] Z$_{\odot}$, 
                  N$_{\rm H}$ = 8.72 [7.06 - 11.1] $\times$ 10$^{22}$ cm$^{-2}$,
                  kT = 5.42 [3.36 - 9.44] keV,  $\chi^2$/dof = 29.0/34,
                  log L$_{\rm X}$ = 31.58 ergs s$^{-1}$.}
\end{deluxetable}

\clearpage

\begin{deluxetable}{llllll}
\tabletypesize{\footnotesize}
\tablewidth{0pc}
\tablecaption{Classical FUors} 
\tablehead{
\colhead{Star}              &
\colhead{Distance}          &
\colhead{A$_{\rm V}$}       &
\colhead{L$_{bol}$}          &
\colhead{log L$_{\rm X}$}   &
\colhead{Refs.}             \\
\colhead{    }              &
\colhead{(pc)}              &
\colhead{(mag)}             &
\colhead{(L$_{\odot}$)}      &
\colhead{(ergs s$^{-1}$)}   &
\colhead{    } 
}
\startdata
FU Ori     & 460             & 1.5 - 2.6   & 340 - 500 & 30.8 $\pm$ 0.4 & 1,2,3,4,5,6,7,8 \nl
V1735 Cyg  & 950 - 1200      & 8.0 - 10.8  & 235       & 31.0 $\pm$ 0.2 & 7,9,10,11      \nl
V1057 Cyg  & 600             & 3.0 - 4.2   & 250 - 800 & $\leq$30.0\tablenotemark{a} & 1,2,3,7,12,13,14,15 \nl
V1515 Cyg  & 1000            & 2.8 - 3.2   & 200       & $\leq$30.5\tablenotemark{b} & 1,2,7,8,13,14,15,16   
\enddata
\tablecomments{
L$_{bol}$ refers to the star$+$disk system and
includes IR-excess emission.
L$_{\rm X}$ (0.5 - 7 keV) is unabsorbed  and the 
uncertainties in L$_{\rm X}$ reflect the range of values
obtained for different emission models (Table 2; SBG06).
Upper limits on unabsorbed L$_{\rm X}$ are from the Portable 
Interactive Mult-Mission Simulator (PIMMS) using EPIC pn 
count rates in the 0.5 - 7 keV range inside
an extraction circle of radius R$_{e}$ = 15$''$
centered on the optical position.  \\
{\em References}.-- (1) SW01; 
                    (2) HK96; 
                    (3) Kenyon et al. 1988;
                    (4) Adams, Lada, \& Shu 1987; 
                    (5) SBG06; 
                    (6) Quanz et al. 2006;
                    (7) this work; 
                    (8) Zhu et al. 2008;
                    (9) H08; (10) Elias 1978; (11) Herbig \& Dahm 2002;
                    (12) Herbig 1977; 
                    (13) Green et al. 2006;
                    (14) Millan-Gabet et al. 2006; 
                    (15) Lorenzetti, D. et al. 2000;
                    (16) Kenyon, Hartmann, \& Kolotilov 1991     
}
\tablenotetext{a}{Based on 23.2 ksec of usable EPIC pn exposure
obtained on 26 Nov 2005 using the medium optical blocking filter
(Observation Id 0302640201). The L$_{\rm X}$ estimate
assumes an underlying two-temperature optically thin thermal 
plasma spectrum with kT$_{1}$ = 0.7 keV, kT$_{2}$ = 3.0 keV,
and log N$_{\rm H}$ = 21.83 cm$^{-2}$.} 
\tablenotetext{b}{Based on 34.2 ksec of usable EPIC pn exposure
obtained on 22 Oct 2006 using the medium optical blocking filter
(Observation Id 0402840101). The L$_{\rm X}$ estimate
assumes an underlying two-temperature optically thin thermal 
plasma spectrum with kT$_{1}$ = 0.7 keV, kT$_{2}$ = 3.0 keV,
and log N$_{\rm H}$ = 21.79 cm$^{-2}$.} 
\end{deluxetable}



\begin{figure}
\figurenum{1}
\epsscale{1.0}
\includegraphics*[width=13.0cm,angle=-90]{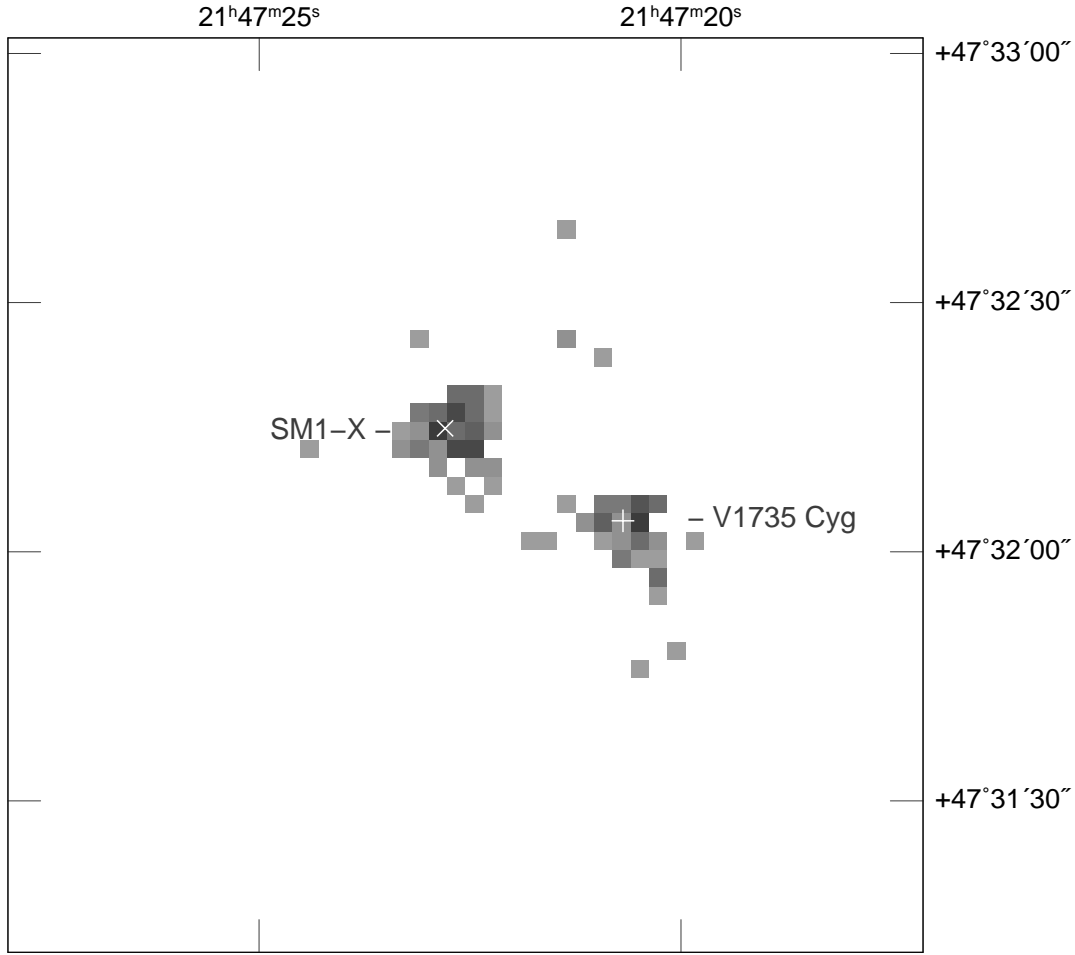}
\caption{
Combined EPIC MOS1 $+$  MOS2 image of the V1735 Cyg region
in the 0.5-7 keV range, rebinned to a pixel size of 2.2$''$. Log
intensity scale. J2000.0 coordinate overlay. The plus sign marks
the {\em HST} GSC position of V1735 Cyg. The $\times$ marks the
{\em Spitzer} position of SM1 (source 34 in H08).
See Table 1 for X-ray source positions.
}

\end{figure}

\clearpage

\begin{figure}
\figurenum{2}
\epsscale{1.0}
\includegraphics*[width=13.0cm,angle=0]{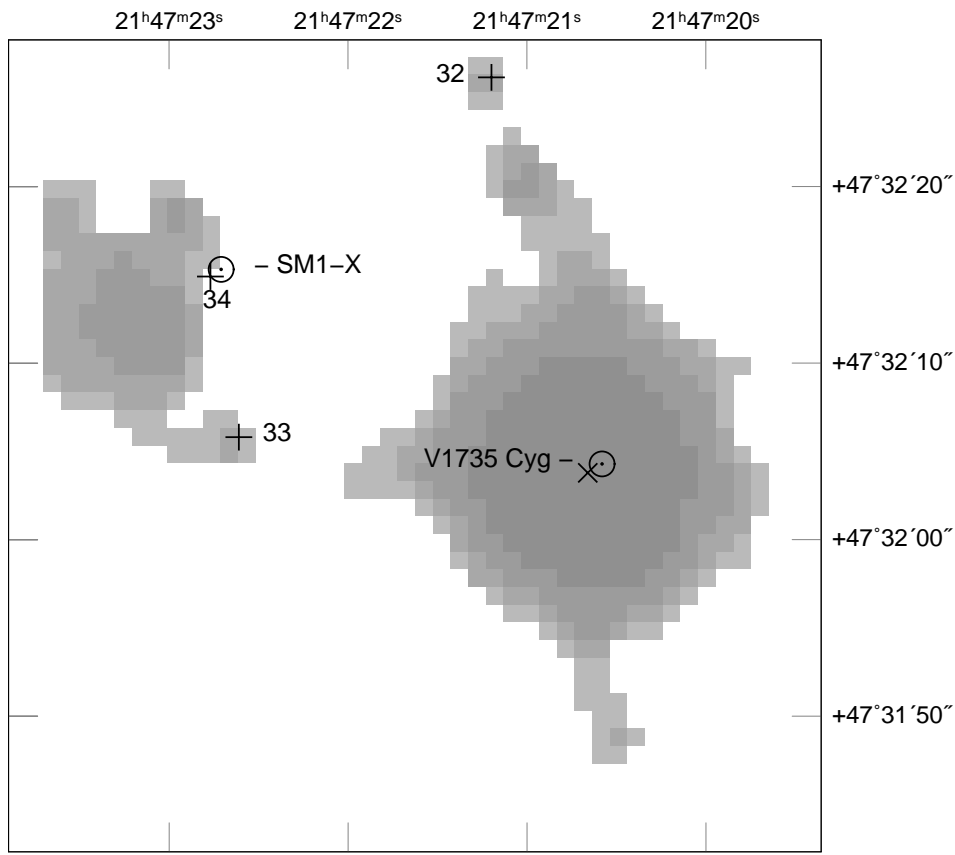}
\caption{
2MASS K$_{s}$ band (2.16 $\mu$m) image of the V1735 Cyg region. 
Circled dots  mark {\em XMM} X-ray  positions of
V1735 Cyg and SM1-X (Table 1). The $\times$ marks the {\em HST} GSC position
of V1735 Cyg. Plus signs ($+$) mark {\em Spitzer} IR-excess sources 
and their corresponding source numbers from Table 5 of H08.
The faint class I {\em Spitzer} sources 32 and 33 are visible 
in the 2MASS image but were undetected by {\em XMM}. Class I
{\em Spitzer} source 34 (SM1) is undetected by 2MASS but is
detected by {\em XMM}. Coordinates are  J2000.0. 
}

\end{figure}

\clearpage

\begin{figure}
\figurenum{3}
\epsscale{1.0}
\includegraphics*[width=13.0cm,angle=-90]{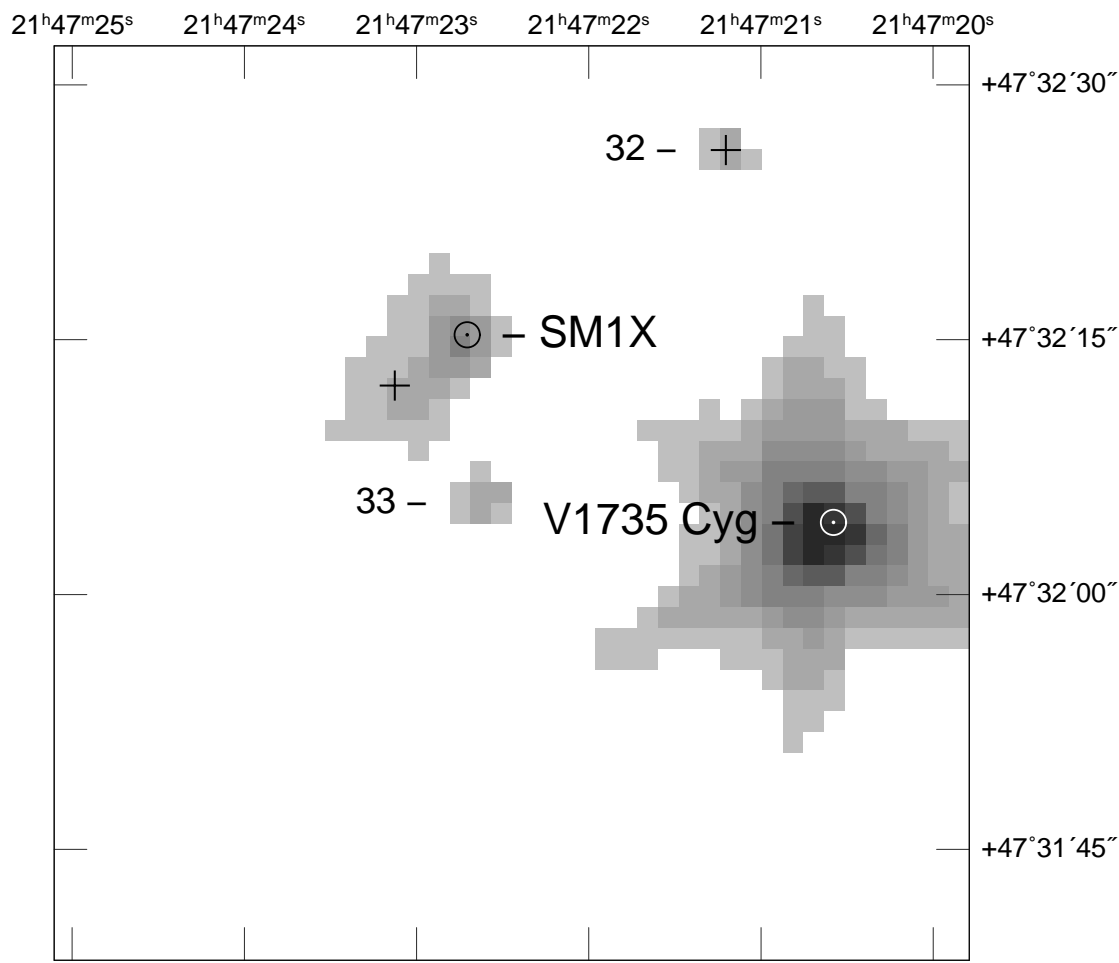}
\caption{
{\em Spitzer} IRAC 3.6 $\mu$m  image of the V1735 Cyg region
from the  {\em Spitzer} archive (Program  94). The IRAC image is 
a mosaic constructed from short 1.2 s frame-time
exposures obtained in the high dynamic range (HDR) mode observation. 
Circles mark {\em XMM} X-ray source positions. Coordinates are  J2000.0.
The X-ray positions lie within 1$''$ of the IRAC peak positions, 
consistent with positional uncertainties. Crosses mark 
2MASS source positions. Objects 32 and 33 were classified as class I
by H08 but were undetected in X-rays.
}

\end{figure}

\clearpage

\begin{figure}
\figurenum{4}
\epsscale{1.0}
\plottwo{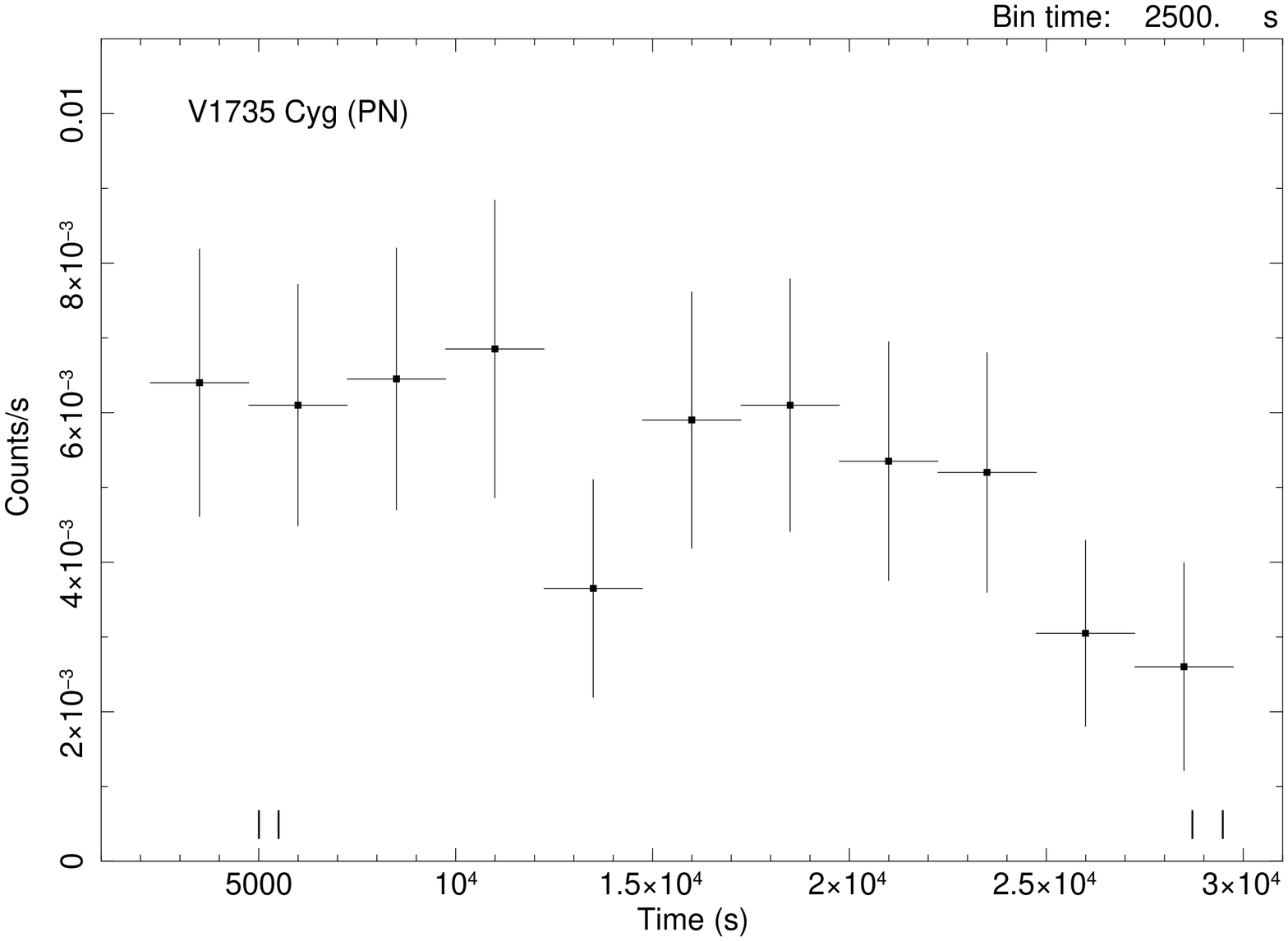}{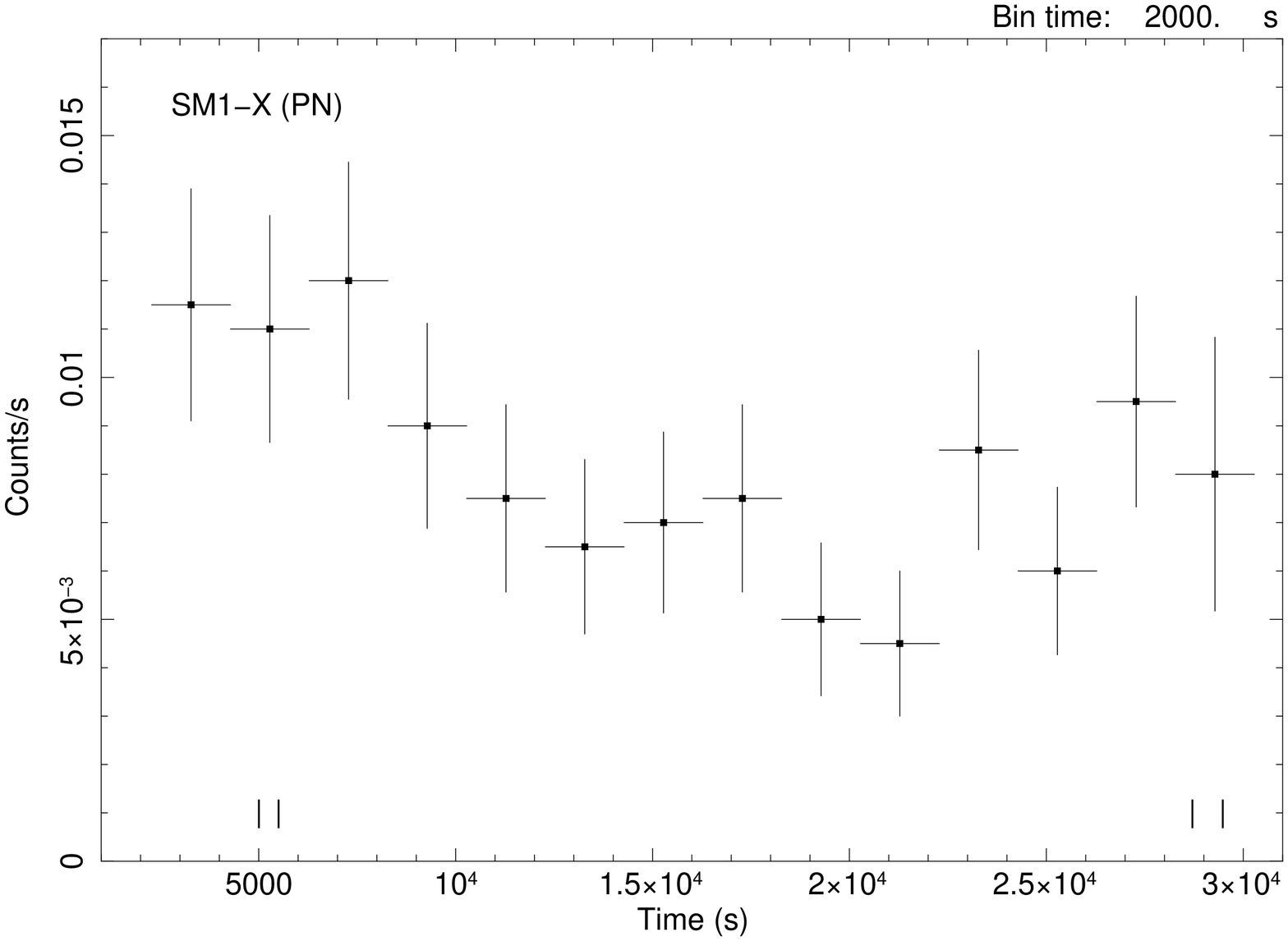}
\caption{
Background-subtracted EPIC PN X-ray light curves of 
V1735 Cyg (2500 s bins) and SM1-X (2000 s bins) in the 0.5 - 7 keV 
range.  The vertical lines on the time axis near 5000 s and 29,000 s
mark short high-background intervals that were excluded from
spectral analysis. Error bars are 1$\sigma$.
}

\end{figure}

\clearpage

\begin{figure}
\figurenum{5}
\epsscale{1.0}
\includegraphics*[width=12cm,angle=-90]{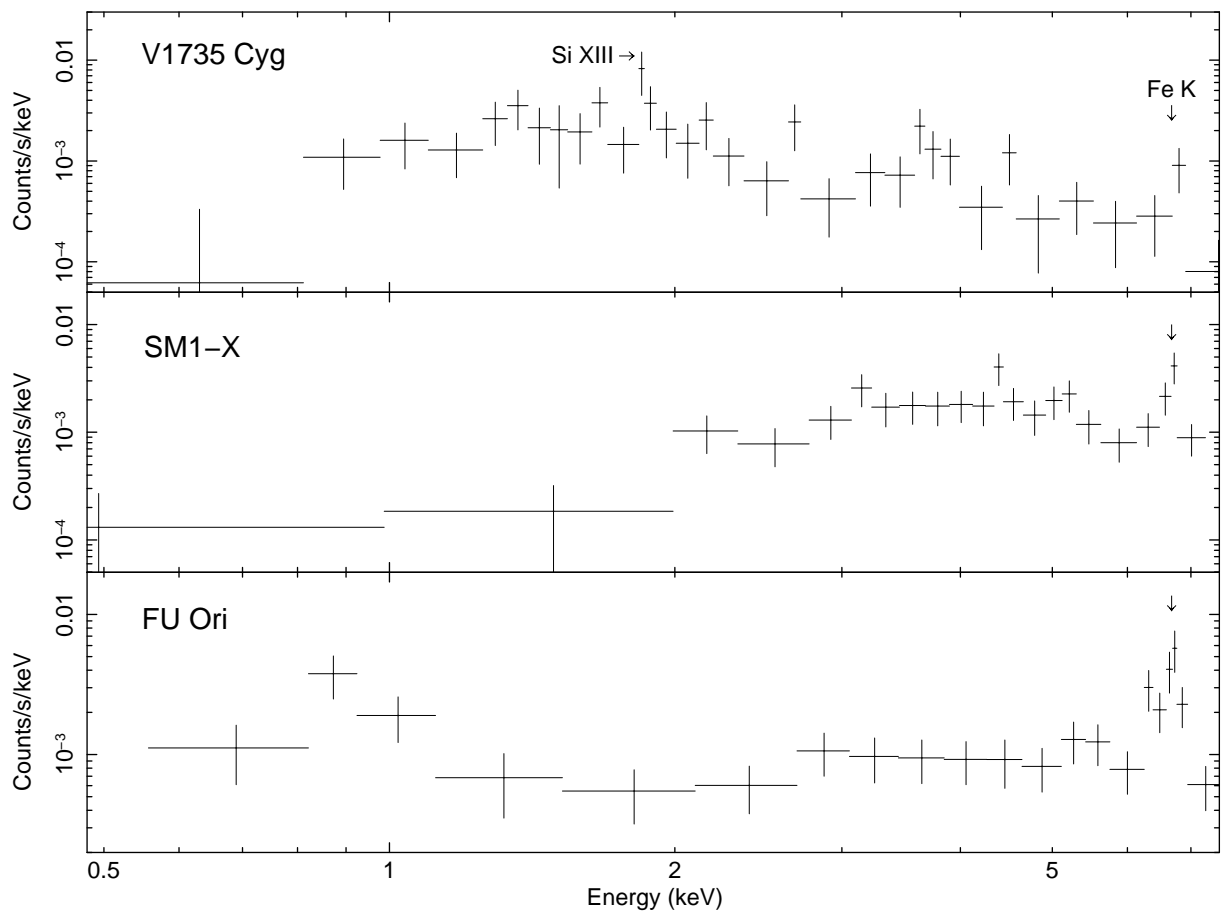}
\caption{
Background-subtracted EPIC PN  spectra  of V1735 Cyg (180 net counts),
SM1-X = XMM J21472271$+$4732153 (247 net counts), and FU Ori
(200 net counts; SBG06). The V1735 Cyg spectrum is grouped to a minimum of
5 counts per bin and the SM1-X and FU Ori spectra are grouped to
a minimum of 10 counts per bin. The identified lines are
Si XIII (1.86 keV) and the Fe K$\alpha$ line complex (6.67 keV). 
}
\end{figure}

\clearpage

\begin{figure}
\figurenum{6}
\epsscale{1.0}
\includegraphics*[width=12cm,angle=-90]{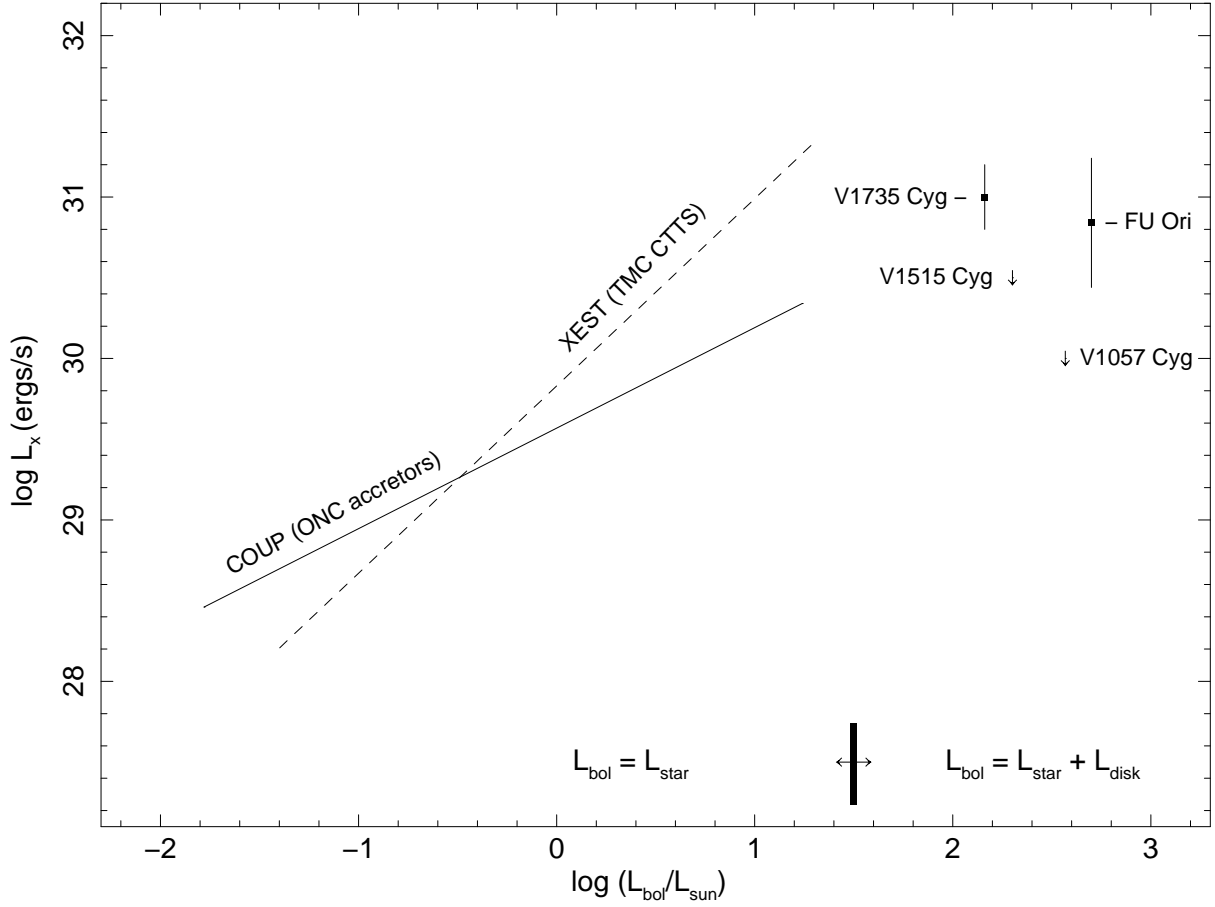}
\caption{
X-ray luminosity (L$_{\rm X}$) versus bolometric luminosity
(L$_{bol}$). For the four FUors, L$_{\rm X}$ is from Table 3
and downward arrows denote upper limits. For the 
FUors,  L$_{bol}$ = L$_{star}$ $+$ L$_{disk}$ and the
disk contribution is  substantial. 
The L$_{bol}$ values for FUors are from SW01, 
except for V1735 Cyg which is from H08. 
For comparison, the regression fits for accreting
sources in the {\em Chandra} COUP survey of the
ONC (solid line; P05) and classical TTS in the 
{\em XMM} XEST survey of the Taurus molecular cloud
(dashed line; T07) are shown. For the COUP and
XEST stars, L$_{bol}$ = L$_{star}$; that is, 
the disk contribution is {\em not} included in  L$_{bol}$.
For  more detailed plots showing the data on which the
COUP and XEST regression lines are based, see
Figure 17 of P05 and Figure 5 of T07.
}
\end{figure}

\clearpage

\end{document}